\documentclass[a4paper, 10pt, conference]{IEEEtran}
\IEEEoverridecommandlockouts
\usepackage{cite}
\usepackage{amsmath,amssymb,amsfonts}
\usepackage{algorithmic}
\usepackage{graphicx}
\usepackage{textcomp}
\usepackage{xcolor}
\def\BibTeX{{\rm B\kern-.05em{\sc i\kern-.025em b}\kern-.08em
    T\kern-.1667em\lower.7ex\hbox{E}\kern-.125emX}}

\newtheorem{definition}{\bf Definition}[section]
\usepackage{mathtools}
\usepackage{url}

\usepackage{listings}
\usepackage{tikz}
\usepackage{environ}
\usepackage{pgfplots}

\begin{document}

\title{Programming Data Structures for Large-Scale Desktop Simulations of Complex Systems*\\
\thanks{This work was supported by the Hasler Foundation under grant No.\ 21017.}
}

\author{\IEEEauthorblockN{Patrik Christen}
\IEEEauthorblockA{\textit{Institute for Information Systems} \\
\textit{FHNW}\\
Olten, Switzerland \\
patrik.christen@fhnw.ch}
}

\maketitle

\begin{abstract}
The investigation of complex systems requires running large-scale simulations over many temporal iterations. It is therefore important to provide efficient implementations. The present study borrows philosophical concepts from Gilbert Simondon to identify data structures and algorithms that have the biggest impact on running time and memory usage. These are the entity $e$-tuple $\mathcal{E}$ and the intertwined update function $\phi$. Focusing on implementing data structures in C\#, $\mathcal{E}$ is implemented as a list of objects according to current software engineering practice and as an array of pointers according to theoretical considerations. Cellular automaton simulations with $10^9$ entities over one iteration reveal that the object-list with dynamic typing and multi-state readiness has a drastic effect on running time and memory usage, especially dynamic typing as it has a big impact on the evolution time. Pointer-arrays are possible to implement in C\# and are more running time and memory efficient as compared to the object-list implementation, however, they are cumbersome to implement. In conclusion, avoiding dynamic typing in object-list based implementations or using pointer-arrays gives evolution times that are acceptable in practice, even on desktop computers.
\end{abstract}


\section{Introduction}

Studying complex systems is as fascinating as it is crucial. Fascinating because new and beautiful things can appear out of nowhere such as the intricate pattern generated by a cellular automaton updating its states according to Wolfram's rule 110 \cite{wolfram.2002}. And crucial because we face enormous problems such as climate change and socio-economic instability, where a complex systems perspective and methodology promise a potential way forward \cite{thurner.2020}. Such complex systems are typically characterised by a large number of elements or entities that interact with each other over time, where the interactions are specific to the entities and change over time \cite{thurner.2018}. The co-evolution of entity states and entity interactions makes complex systems difficult if not impossible to treat analytically, since this would lead to a system of dynamical equations dynamically coupled to their boundary conditions \cite{thurner.2018}. A more natural way to determine the dynamics of a complex system is to specify an algorithm consisting of rules on how entities and their interactions update over time \cite{thurner.2018}, which is nicely illustrated with cellular automata. The algorithmic approach allows us to simulate complex systems and run experiments for hypothesis testing. In contrast to the analytical approach where equations are solved and results can be directly calculated for any point in time, the algorithmic approach requires the calculation of entity states in every time iteration over the simulated time. It is not possible to jump ahead in time, which makes this approach computationally expensive, especially because complex systems are usually composed of a large number of entities and a series of experiments are run to study their behaviour in response to certain changes.

The programming of efficient algorithms and data structures for modelling complex systems is therefore essential. There is, however, no standardised method for modelling complex systems which makes this task a challenge. We recently developed the so-called allagmatic method \cite{DelFabbro.arXiv.Adaptation,Christen.arXiv.Formalism,Christen.SMC2019}, which is a general approach for modelling and simulating complex systems including the definition of data structures and algorithms. It is inspired by philosophical concepts such as Gilbert Simondon's structure and operation \cite{DelFabbro.2021,Simondon.2020}, and Alfred North Whitehead's adaptation and control \cite{Debaise.2017,Whitehead.1978}. These concepts are implemented as a system metamodel with program code that not only allows the concepts to be defined precisely, but also allows them to be executed or run in relation to other concepts. The allagmatic method proved to be useful to generate concrete computer models such as cellular automata and artificial neural networks \cite{DelFabbro.arXiv.Adaptation,Christen.SMC2019} as well as to guide automatic programming by providing abstract model building blocks from which systems are composed of \cite{Christen.IWIBook}. These model building blocks can be put together in an automatic way by concretising them into more and more concrete models of complex systems. With self-modifying code, on the other hand, it allows controlled modification of specific code regions \cite{Christen.arXiv.Curb,Christen.arXiv.OEE,Christen.OEE4}.

It is common to use C++ for efficient implementations and to run the code on a supercomputer. The allagmatic method was first developed in C++ using classes and generic programming (template metaprogramming) \cite{Alexandrescu.2001} to describe the system metamodel \cite{Christen.SMC2019}. In later studies on self-modifying code, the system metamodel was reimplemented in C\# \cite{Christen.arXiv.Curb,Christen.arXiv.OEE,Christen.OEE4}, because it allows compiling and running code as an object, provides reflection capabilities, and generally comes with a vast API. Compiling and running code as an object in C\# is faster than writing code into a file, call the system to compile and then run the executable in C++. Reflection is rather limited if not missing in C++ while C\# provides respective classes. Although it is possible, self-modifying code and reflection are not welcome concepts to be implemented on supercomputers, where you want to have full control over the code and C\# might only run in cloud systems and not on supercomputers. Both, cloud systems and supercomputers, are usually designed with nodes that consist of little memory and one is required to parallelise the code. Since complex systems are composed of a large number of elements, this can be an issue, especially if one wants to create prototypes and not invest too much time into parallel implementations. Desktop computers have become fast too and, most relevant for this application, they provide large memory. In addition, desktop computers are also an interesting alternative when it comes to sensitive data allowing computation on-site such as in a hospital or in a military application without any data transfer over the internet.

The allagmatic method was developed with a focus on modelling and simulating complex systems according to philosophical concepts, however, the respective data structures and algorithms were not optimally implemented with respect to running time and memory usage. In the present study, an implementation based on objects and another based on pointers are described and compared in terms of running time and memory usage on a desktop computer. In the first section of the paper, I introduce the allagmatic method and then use it to abstractly define data structures and algorithms of complex systems in the second section. The third section deals with the programming and thus specific implementation of the defined data structures and algorithms of the second section. The fourth section describes large-scale experiments and presents the results of their running time and memory usage on a desktop computer. Finally, the differences in performance are discussed and an outlook on some of the next steps is provided.

\section{The Allagmatic Method}

The French philosopher Gilbert Simondon describes objects and processes (or, more generally, systems) in terms of structures and operations \cite{DelFabbro.2021,Simondon.2020}. Structures capture the static or spatial side of a system while operations capture their dynamic or temporal side. Thereby, Simondon emphasises that structures and operations are tightly intertwined and should therefore not be studied in isolation. This means that every system can be described from a structural and operational perspective and it also means that each operation must have a respective structure to operate on and vice versa. The philosophy of Alfred North Whitehead is compatible with Simondon's system perspective and adds to it important concepts such as entity, adaptation, and control \cite{Debaise.2017,Whitehead.1978}. Whitehead describes systems in terms of entities that constantly interact with each other forming higher level systems (or in his terminology societies and nex\={u}s). This allows him to clearly define deep concepts including adaptation, control, and complexity.

The so-called allagmatic method was initially developed based on Simondon's structure and operation \cite{Christen.SMC2019}, and later extended with Whitehead's adaptation and control \cite{DelFabbro.arXiv.Adaptation,Christen.arXiv.Formalism}. These concepts enable the modelling and simulation of systems and especially of complex systems. The method consists of a system metamodel that describes the modelled system at the most abstract level in the so-called virtual regime. At this point the system is virtual and defined in terms of its abstract structures and operations. These structures and operations are then concretised in the so-called metastable regime. Here, details such as the size of the system and specific operations are defined. Once a concrete model of a system is created, it can be executed or run in the so-called actual regime. This is where the simulation is run. There are also feedback loops between the regimes describing higher level concepts such as adaptation. In adaptation, the method cycles between the metastable and actual regimes \cite{DelFabbro.arXiv.Adaptation,Christen.arXiv.Formalism}.

The system metamodel of the allagmatic method describes a (complex) system as a network of entities that change their states over time in response to certain update rules and states of connected entities in the network, here called the milieu. Adaptation means changing these update rules or milieus. We can formally define a system model $\mathcal{SM}$ (which also consists of the system metamodel) with a tuple of structures $S$ and operations $O$ specific to complex systems as follows:

\begin{definition}
	\begin{equation}
		\mathcal{SM} \coloneqq (\mathcal{E},Q,\mathcal{M},\mathcal{U},\mathcal{A},\mathcal{P},\dots,\hat{s}_s,\phi,\psi,\dots,\hat{o}_o),
	\end{equation}
where $\mathcal{E}$ is an entity $e$-tuple describing the entity states, $Q$ is a set describing the possible entity states, $\mathcal{M}$ is a milieu $e$-tuple describing the connected entities for each entity, $\mathcal{U}$ is an update rules $u$-tuple describing the dynamics of entity states, $\mathcal{A}$ is an adaptation rule $a$-tuple describing the adaptation of the update rules $\mathcal{U}$ taking into account the adaptation end described with the $p$-tuple $\mathcal{P}$, $\hat{s}_i$ are any further structures, $\phi$ is the update function executing the update rules $\mathcal{U}$ on the entity states $\mathcal{E}$, $\psi$ is the adaptation function executing the adaptation rules $\mathcal{A}$ on the update rules $\mathcal{U}$ and milieus $\mathcal{M}$, and $\hat{o}_j$ are any further operations.
\end{definition}

\section{Data Structures of Complex Systems}

We now use the allagmatic method to define the essential data structures of complex systems. Simondon's structures can be seen as data structures and operations as algorithms in the way they are classically described in textbooks on algorithms and data structures \cite{Cormen.2022}. It is interesting to note at this point that also in the field of algorithms and data structures one should study the two not in isolation but in relation to each other. So Simondon has already foreseen this.

If we are guided by the allagmatic method, we can see that at the core there are the entity states $\mathcal{E}$ updated by the update function $\phi$ and it is mostly them that determine the running time and memory usage. The milieu $\mathcal{M}$ is also important as it determines the network structure of the system. The other structures and operations are of course important as well, however, not with respect to running time and memory usage because these structures are rather small and these operations are executed on the small structures and only a few times. In contrast, the entity state $e$-tuple $\mathcal{E}$ holds the states of every entity and has a size of $e$, which is generally large in complex systems. The update function $\phi$ operates on every entity and updates its state in every iteration over time.

Further, we can apply the theory of algorithms and data structures \cite{Cormen.2022} to this core data structure and algorithm. The network structure described with $\mathcal{E}$ and $\mathcal{M}$ can be well captured with a graph, and graphs, on the other hand, can be well represented by adjacency lists or adjacency matrices. Since in our case the total number of entities $e$ is much bigger than the number of entities in a milieu of a particular entity $m$, an adjacency list is favourable in terms of space-efficiency. Each entity is updated by $\phi$ and states of the entities in the milieu can be easily accessed via index. If the milieu $\mathcal{M}$ changes, the references have to be updated as well. In case a new entity is added to the system, one does not have to rearrange $\mathcal{E}$ and $\mathcal{M}$, it can be added to the end of the adjacency list. It is therefore favourable in terms of time-efficiency to base the adjacency list on an array of arrays. Thereby, the array represents the index of $\mathcal{E}$ and the arrays within it represent $\mathcal{M}$.

\section{Programming of Data Structures}

The adjacency list and the update function operating on it were implemented here in two different ways: first, according to standard software engineering practice and second, an optimised version with the theory of algorithms and data structures in mind. Both versions were implemented in C\# bearing in mind that most applications will require concepts such as reflection that would be difficult or cumbersome to implement in C++. Please consult Skeet's book \textit{C\# in Depth} \cite{Skeet.2019} or the .NET Documentation \cite{DotNETDocu} for more details on the capabilities of C\#.

 The first version of the implementation is in accordance with standard software engineering practice. In the case of C\#, this means object-oriented programming and making use of data containers provided by the API. An entity is implemented with a class \texttt{Entity} consisting of three fields: first, a field storing the state of the entity. Second, a field storing the updated state of the entity. This is required because one cannot overwrite the state directly during the update since the current state might still be used by another entity. And third, a field storing the milieu. The first two fields are of the same data type, in the example listing below they are of type dynamic. This data type was chosen because according to the allagmatic method, entity states are concretised. In addition, boolean type was also used for comparison as well as considering single-state and multi-state entities, each of them with boolean type and dynamic type. Multi-state means that instead of a primitive data type, a list has been used to allow storing of multiple states per entity. Only one state was implemented, thus it is rather multi-state ready. The milieu is an array of references to entity objects. In C\#, arrays of objects are provided by the \texttt{List} class \cite{ListClass}. It is an array and should not be confused with a list such as a linked list.

\begin{lstlisting}[basicstyle=\small]
public class Entity
{
  private dynamic state;
  private dynamic nextState;
  private List<Entity> milieu;
}
\end{lstlisting}

In the main program, another array is created to store the entities. The \texttt{List} class is used since objects are stored in an array. The array is first declared and then objects are added to it with the method \texttt{Add} in a first for-loop and the milieu is generated with the method \texttt{GenerateMilieu} in a second for-loop as shown in the listing below. First and last entities in the array are treated separately outside the for-loop defining certain boundary conditions. It implements the structures $\mathcal{E}$ and $\mathcal{M}$.

\begin{lstlisting}[basicstyle=\small]
List<Entity> entities = new List<Entity>();
for (int i=0; i<numberOfEntities; i++)
{
  Entity entity = new Entity();
  entities.Add(entity);
}
for (int i=1; i<numberOfEntities-1; i++)
{
  entities[i].GenerateMilieu(entities, i);
}
\end{lstlisting}

The \texttt{Entity} class also consists of a method implementing the update function $\phi$. It updates the states of each entity in each time iteration according to predefined update rules implemented with if-statements in a first for-loop. Once the states of the next iteration are determined and written to the field \texttt{nextState}, a second for-loop is used to overwrite the field value of \texttt{state} with the field value of \texttt{nextState} for each entity. This is done in every iteration over time in an outer for-loop. The listing below shows the three for-loops in the main program.

\begin{lstlisting}[basicstyle=\small]
for (int i=0; i<numberOfUpdateIterations; i++)
{
  for (int j=0; j<numberOfEntities; j++)
  {
    entities[j].UpdateFunction();
  }
  for (int j=0; j<numberOfEntities; j++)
  {
    entities[j].UpdateFields();
  }
}
\end{lstlisting}

The second version of the implementation is optimised according to the theory of algorithms and data structures and specific to C\#. That means this version tries to directly implement the theoretical concepts, i.e. references and arrays, and it tries to avoid programming concepts known to increase running time or memory usage. From theory \cite{Cormen.2022} we already know from the first version that an adjacency list implemented with an array of references pointing to other arrays is most suitable for the update function. In the first version, this was implemented with an array (specifically a list in C\#) of references pointing to objects, and these objects aggregate entity states and milieus. The second implementation is closer to the idea of having an array of references to other arrays by implementing basic arrays and pointers. It does not involve objects but actual pointers as used in C++. It thus avoids objects, which take time to set up by the constructor and maintain in memory, especially memory allocation and garbage collection. This is also the reason why it has been suggested to reuse allocated memory rather than newly allocate it.

The entity states and thus $\mathcal{E}$ are directly stored in a basic array as shown in the listing below. By basic array, I mean an array of primitive data types, without dynamic resizing, and without dynamic data type. It is not to be confused with the \texttt{List} class that provides an array that can hold objects and dynamically adapt its size. Note that both, dynamic resizing and dynamic data type, are not possible to implement with pointers in C\#. There are two arrays \texttt{entity\_state} and \texttt{entity\_nextState} in the main program holding the current and next states for each entity. A two-dimensional basic array \cite{TwoDimensionalArray} is used for the multi-state ready implementation. And there is an array of arrays \texttt{milieu} to describe the milieu $\mathcal{M}$. Also the third array is a basic array, however, it is a pointer array and a so-called jagged array \cite{JaggedArrays}. A jagged array is an array of arrays, where the contained arrays can be of different sizes.

\begin{lstlisting}[basicstyle=\small]
bool[] entity_state = 
  new bool[numberOfEntities];
bool[] entity_nextState = 
  new bool[numberOfEntities];
bool*[][] milieu = 
  new bool*[numberOfEntities][];
\end{lstlisting}

While creating and using pointers in C++ is straightforward, it is not so in languages with automatic memory management such as in C\#. Memory allocated in C++ stays at the physical memory location and thus also at the address as long as the programmer does not instruct the program otherwise. Memory allocated in C\#, in contrast, might be optimised at runtime, which also includes changing the physical memory location and thus the address. Pointers are variables that hold memory addresses and thus rely on allocated memory to stay in the physical memory location. C\# provides the \texttt{fixed} statement \cite{FixedStatement} to pin or fix the allocated memory to its location. It prevents the garbage collector from moving allocated memory to another location \cite{FixedStatement}. Pointers in C\# are only possible with the \texttt{fixed} statement and within an \texttt{unsafe} context \cite{UnsafeContext}. Here, the main method is defined in unsafe context to make it possible to use pointers. Pointers are used in the \texttt{milieu} array. The milieu of an entity is implemented by pointing to the neighbouring entities in the entity states array. This means that the memory address of individual elements of the entity states array \texttt{entity\_state} are stored in the \texttt{milieu} array. For each entity, a pointer is created for each of its neighbours in a \texttt{fixed} statement, and then a pointer array is created to store these pointers as shown in the follwoing listing.

\begin{lstlisting}[basicstyle=\small]
for (int i=1; i<numberOfEntities-1; i++)
{
  fixed (bool* pointerToNeighbour1 = 
    &entity_state[i-1])
  {
    fixed (bool* pointerToNeighbour2 = 
      &entity_state[i+1])
    {
      milieu[i] = new bool*[2];
      milieu[i][0] = pointerToNeighbour1;
      milieu[i][1] = pointerToNeighbour2;
    }
  }
}
\end{lstlisting}

While the use of objects is avoided, it is still possible to use methods. To run the update function $\phi$, a static method was implemented in the \texttt{Entity} class as shown in the listing below. It updates entity states according to predefined update rules implemented in the same way as in the first implementation. Besides the method being static, it has to be defined unsafe to use pointers as parameters. Also in this case, entity states are first determined and stored in \texttt{entity\_nextState} for each entity in a for-loop. Once determined, another for-loop is used through all the entities to overwrite the values in \texttt{entity\_state} with the new value stored in \texttt{entity\_nextState}.

\begin{lstlisting}[basicstyle=\small]
for (int i=0; i<numberOfUpdateIterations; i++)
{
  Entity.PointerArrayUpdateFunction
    (entity_state, entity_nextState, milieu);
}
\end{lstlisting}

\section{Large-Scale Experiments and Results}

The object-list and pointer-array based implementations are compared by running a large-scale experiment with a system size of $10^9$ entities. Running time is measured using the \texttt{Stopwatch} class \cite{StopwatchClass} and memory usage is measured using the \texttt{top} command of the Terminal application of MacOS X. Experiments are run on a desktop computer (Apple Mac Pro) simulating a large elementary cellular automaton with Wolfram's rule 30 \cite{wolfram.2002}. Initially, the state of each entity is set false, with the exception of the array element at index $10^9/2$, which is set true. The different implementations are compared by measuring the running time to set up the arrays, which is here called development time, and to run one iteration of the cellular automaton, which is here called evolution time. Memory usage is given as the maximum memory assigned to the process running the program.

Before running the large-scale simulation of the cellular automaton for the different implementations, they were tested on a much smaller scale of $31$ entities and run for $16$ iterations. The expected pattern for Wolfram's rule 30 emerged (Fig. \ref{fig_verification}), verifying the implementations.

\begin{figure}[tb]
\begin{center}
\texttt{0000000000000001000000000000000
        0000000000000011100000000000000
        0000000000000110010000000000000
        0000000000001101111000000000000
        0000000000011001000100000000000
        0000000000110111101110000000000
        0000000001100100001001000000000
        0000000011011110011111100000000
        0000000110010001110000010000000
        0000001101111011001000111000000
        0000011001000010111101100100000
        0000110111100110100001011110000
        0001100100011100110011010001000
        0011011110110011101110011011100
        0110010000101110001001110010010
        1101111001101001011111001111111}
\end{center}
\caption{Verification output of running the cellular automaton implementations of Wolfram's rule 30 with 31 entities and for 16 update iterations.}
\label{fig_verification}
\end{figure}

Running time and memory usage for $10^9$ entities and one update iteration are shown in Tab. \ref{tab1}. The object-list development time is mostly affected by the combination of dynamic type and multi-state ready implementation. Dynamic type and single-state implementation as well as static typing take much less time. The situation is somewhat different for object-list evolution. There it is clearly the dynamic type that affects the running time the most. In terms of memory usage, it is interesting to note that both, dynamic typing and multi-state readiness, increase memory usage. The pointer-array development time is not affected by multi-state readiness, however the evolution time is doubling the required running time. Dynamic typing is not possible with pointers and thus no comparison can be made. Memory usage is also not affected by multi-state readiness.

\begin{table}[tb]
\caption{Running Time [hh:mm:ss] and Memory Usage [GB]\\for $10^9$ Entities and One Update Iteration}
\begin{center}
\begin{tabular}{|p{1.8cm}|p{1.0cm}|p{1.0cm}|p{1.0cm}|p{1.0cm}|}
\hline
\textbf{}&\multicolumn{2}{|c|}{\textbf{Boolean Type}}&\multicolumn{2}{|c|}{\textbf{Dynamic Type}} \\
\cline{2-5} 
\textbf{} & \textbf{\textit{Single-State}}& \textbf{\textit{Multi-State}}& \textbf{\textit{Single-State}} & \textbf{\textit{Multi-State}} \\
\hline
Object-List Development Time& 00:09:07 & 00:24:49 & 00:17:17 & 03:53:16 \\
\hline
Pointer-Array Development Time& 00:01:27 & 00:01:26 & - & - \\
\hline
Object-List Evolution Time& 00:00:25 & 00:00:58 & 00:42:44 & 01:08:32 \\
\hline
Pointer-Array Evolution Time& 00:00:07 & 00:00:14 & - & - \\
\hline
Object-List Memory Usage& 119 & 249 & 196 & 465 \\
\hline
Pointer-Array Memory Usage& 47 & 47 & - & - \\
\hline
\end{tabular}
\label{tab1}
\end{center}
\end{table}

\section{Discussion and Conclusion}

Data structures and algorithms for simulating complex systems are successfully defined with the allgmatic method \cite{DelFabbro.arXiv.Adaptation,Christen.arXiv.Formalism,Christen.SMC2019}. It provides abstract descriptions and formalisms for structures capturing the spatial dimension and operations capturing the temporal dimension of complex systems according to the philosophy of Gilbert Simondon \cite{DelFabbro.2021,Simondon.2020}. Structures and operations can directly be understood as data structures and algorithms, respectively, which makes Simondon's philosophy relevant in theoretical computer science and programming.

The entities $e$-tuple $\mathcal{E}$ and the intertwined update function $\phi$ are most relevant with respect to running time and memory usage. The present study focuses on the implementation of $\mathcal{E}$ and presents different options for implementing it in C\#. One implementation is according to the current practice in software engineering following object-oriented programming. It implements a list of entity objects, which allows most of the modern features such as dynamic resizing and dynamic typing. This greatly increases running time and memory usage, especially if one combines dynamic typing and lists (Tab. \ref{tab1}). Dynamic typing should be avoided since it not only increases development time, it also substantially increases evolution time, which is most important because many iterations usually need to be run. The pointer-array is the most time and memory efficient implementation although much more cumbersome to implement. Multi-dimensional and jagged arrays are implemented as basic arrays referencing other arrays, which seems to avoid producing any overhead since multi-state readiness does not affect running time or memory usage.

As long as the memory usage is not exceeding the memory capacity of the computer, it is expected to get similar results using different computers but the same operating system. With at least 47 GB memory usage per iteration, the memory capacity can be exceeded in only a few iterations if the simulation results of each iteration are stored. The storing of simulation results remains to be investigated and an efficient algorithm implemented. Running the simulations on a different operating system might lead to different results. Some processes related to memory management could affect development and evolution time.

The current study also shows that modern desktop computers are able to run relatively large systems with $10^9$ entities and over many iterations in time. It can thus be concluded that avoiding dynamic typing in object-list based implementations or using pointer-arrays gives evolution times that are acceptable in practice, even on desktop computers. It can also be concluded that C\# allows the combination of efficient tools like pointers and basic arrays with more convenient and modern tools as well as a rich API.

Many questions, however, remain unanswered, for example, regarding storing simulation results and the update function itself. Future studies might look into parallel file writing and computing as well as other optimisations of storing large data sets and algorithms of the update function.

%\section*{Acknowledgment}
%
%The preferred spelling of the word ``acknowledgment'' in America is without 
%an ``e'' after the ``g''. Avoid the stilted expression ``one of us (R. B. 
%G.) thanks $\ldots$''. Instead, try ``R. B. G. thanks$\ldots$''. Put sponsor 
%acknowledgments in the unnumbered footnote on the first page.

\bibliographystyle{ieeetr}
\bibliography{CS_Data_Structure_SMC}

\begin{thebibliography}{10}

\bibitem{wolfram.2002}
S.~Wolfram, {\em A New Kind of Science}.
\newblock Champaign, IL: Wolfram Media, 2002.

\bibitem{thurner.2020}
S.~Thurner, {\em Die Zerbrechlichkeit der Welt}.
\newblock Wien: edition a, 2020.

\bibitem{thurner.2018}
S.~Thurner, R.~Hanel, and P.~Klimek, {\em Introduction to the Theory of Complex
  Systems}.
\newblock New York, NY: Oxford University Press, 2018.

\bibitem{DelFabbro.arXiv.Adaptation}
O.~Del~Fabbro and P.~Christen, ``{Philosophy-Guided Modelling and
  Implementation of Adaptation and Control in Complex Systems},'' in {\em IEEE
  World Congress On Computational Intelligence (IEEE WCCI)}, 2022.
\newblock arXiv:2009.00110 [cs.NE].

\bibitem{Christen.arXiv.Formalism}
P.~Christen and O.~Del~Fabbro, ``{Philosophy-Guided Mathematical Formalism for
  Complex Systems Modelling},'' in {\em 2022 IEEE International Conference on
  Systems, Man and Cybernetics (IEEE SMC)}, 2022.
\newblock arXiv:2005.01192 [cs.NE].

\bibitem{Christen.SMC2019}
P.~Christen and O.~Del~Fabbro, ``{Cybernetical Concepts for Cellular Automaton
  and Artificial Neural Network Modelling and Implementation},'' in {\em 2019
  IEEE International Conference on Systems, Man and Cybernetics (IEEE SMC)},
  pp.~4124--4130, 2019.
\newblock arXiv:2001.02037 [cs.OH].

\bibitem{DelFabbro.2021}
O.~Del~Fabbro, {\em Philosophieren mit Objekten: Gilbert Simondons prozessuale
  Individuationsontologie}.
\newblock Frankfurt and New York: Campus Verlag, 2021.

\bibitem{Simondon.2020}
G.~Simondon, {\em {Individuation in Light of Notions of Form and Information
  (T. Adkins, trans.)}}, vol.~I \& II.
\newblock Minneapolis: University of Minnesota Press, 2020.

\bibitem{Debaise.2017}
D.~Debaise, {\em Nature as Event: The Lure of the Possible (M. Halewood,
  trans.)}.
\newblock Durham and London: Duke University Press, 2017.

\bibitem{Whitehead.1978}
A.~N. Whitehead, {\em {Process and Reality: An Essay in Cosmology (D. R. Grifin
  and D. W. Sherburne, eds.)}}.
\newblock New York, NY: Free Press, corrected~ed., 1978.

\bibitem{Christen.IWIBook}
P.~Christen and O.~Del~Fabbro, ``Automatic programming of cellular automata and
  artificial neural networks guided by philosophy,'' in {\em New Trends in
  Business Information Systems and Technology} (R.~Dornberger, ed.), vol.~294
  of {\em Studies in Systems, Decision and Control}, pp.~131--146, Cham:
  Springer, 2021.
\newblock arXiv:1905.04232 [cs.AI].

\bibitem{Christen.arXiv.Curb}
P.~Christen, ``{Curb Your Self-Modifying Code},'' in {\em 2022 IEEE
  International Conference on Systems, Man and Cybernetics (IEEE SMC)}, 2022.
\newblock arXiv:2202.13830 [cs.SE].

\bibitem{Christen.arXiv.OEE}
P.~Christen, ``{Self-Modifying Code in Open-Ended Evolutionary Systems},''
  2022.
\newblock arXiv:2201.06858 [cs.NE].

\bibitem{Christen.OEE4}
P.~Christen, ``{Modelling and Implementing Open-Ended Evolutionary Systems},''
  in {\em The Fourth Workshop on Open-Ended Evolution (OEE4), The 2021
  Conference on Artificial Life (ALife)}, 2021.
\newblock arXiv:2201.06858v1 [cs.NE].

\bibitem{Alexandrescu.2001}
A.~Alexandrescu, {\em Modern C++ Design: Generic Programming and Design
  Patterns Applied}.
\newblock Upper Saddle River, NJ: Addison-Wesley, 2001.

\bibitem{Cormen.2022}
T.~H. Cormen, C.~E. Leiserson, R.~L. Rivest, and C.~Stein, {\em Introduction to
  Algorithms}.
\newblock Cambridge, MA: MIT Press, 4th~ed., 2022.

\bibitem{Skeet.2019}
J.~Skeet, {\em C\# in Depth}.
\newblock Shelter Island, NY: Manning, 2019.

\bibitem{DotNETDocu}
{The .NET C\# Documentation}, ``{C\# documentation}.''
  \url{https://docs.microsoft.com/en-gb/dotnet/csharp/}, 2022.
\newblock [Accessed 19 April 2022].

\bibitem{ListClass}
{The .NET C\# Documentation}, ``{List$<$T$>$ Class}.''
  \url{https://docs.microsoft.com/en-gb/dotnet/api/system.collections.generic.list-1?view=net-6.0},
  2022.
\newblock [Accessed 19 April 2022].

\bibitem{TwoDimensionalArray}
{The .NET C\# Documentation}, ``{Multidimensional Arrays (C\# Programming
  Guide)}.''
  \url{https://docs.microsoft.com/en-gb/dotnet/csharp/programming-guide/arrays/multidimensional-arrays},
  2022.
\newblock [Accessed 25 April 2022].

\bibitem{JaggedArrays}
{The .NET C\# Documentation}, ``{Jagged Arrays (C\# Programming Guide)}.''
  \url{https://docs.microsoft.com/en-gb/dotnet/csharp/programming-guide/arrays/jagged-arrays},
  2022.
\newblock [Accessed 20 April 2022].

\bibitem{FixedStatement}
{The .NET C\# Documentation}, ``{fixed Statement (C\# Reference)}.''
  \url{https://docs.microsoft.com/en-gb/dotnet/csharp/language-reference/keywords/fixed-statement},
  2022.
\newblock [Accessed 20 April 2022].

\bibitem{UnsafeContext}
{The .NET C\# Documentation}, ``{unsafe (C\# Reference)}.''
  \url{https://docs.microsoft.com/en-gb/dotnet/csharp/language-reference/keywords/unsafe},
  2022.
\newblock [Accessed 20 April 2022].

\bibitem{StopwatchClass}
{The .NET C\# Documentation}, ``{Stopwatch Class}.''
  \url{https://docs.microsoft.com/en-gb/dotnet/api/system.diagnostics.stopwatch?view=net-6.0},
  2022.
\newblock [Accessed 20 April 2022].

\end{thebibliography}

%\begin{thebibliography}{00}
%\bibitem{b1} G. Eason, B. Noble, and I. N. Sneddon, ``On certain integrals of Lipschitz-Hankel type involving products of Bessel functions,'' Phil. Trans. Roy. Soc. London, vol. A247, pp. 529--551, April 1955.
%\bibitem{b2} J. Clerk Maxwell, A Treatise on Electricity and Magnetism, 3rd ed., vol. 2. Oxford: Clarendon, 1892, pp.68--73.
%\bibitem{b3} I. S. Jacobs and C. P. Bean, ``Fine particles, thin films and exchange anisotropy,'' in Magnetism, vol. III, G. T. Rado and H. Suhl, Eds. New York: Academic, 1963, pp. 271--350.
%\bibitem{b4} K. Elissa, ``Title of paper if known,'' unpublished.
%\bibitem{b5} R. Nicole, ``Title of paper with only first word capitalized,'' J. Name Stand. Abbrev., in press.
%\bibitem{b6} Y. Yorozu, M. Hirano, K. Oka, and Y. Tagawa, ``Electron spectroscopy studies on magneto-optical media and plastic substrate interface,'' IEEE Transl. J. Magn. Japan, vol. 2, pp. 740--741, August 1987 [Digests 9th Annual Conf. Magnetics Japan, p. 301, 1982].
%\bibitem{b7} M. Young, The Technical Writer's Handbook. Mill Valley, CA: University Science, 1989.
%\end{thebibliography}

\end{document}